\documentclass[preprint2]{aastex6}
\usepackage{amsmath}

\newcommand{\abar}{\bar{A}}
\newcommand{\zbar}{\bar{Z}}
\newcommand{\chiT}{\chi_{T}}
\newcommand{\chir}{\chi_{\rho}}
\newcommand{\Ye}{Y_{\mathrm{e}}}

\newcommand{\dlnp}{d \ln P}
\newcommand{\dlna}{d \ln \abar}
\newcommand{\dlnz}{d \ln \zbar}
\newcommand{\dlnye}{d \ln \Ye}

\newcommand{\ptot}{P_{\mathrm{tot}}}
\newcommand{\pele}{P_{\mathrm{ele}}}
\newcommand{\pion}{P_{\mathrm{ion}}}
\newcommand{\pcou}{P_{\mathrm{cou}}}

\newcommand{\grada}{\nabla_{\mathrm{ad}}}
\newcommand{\gradT}{\nabla_{T}}

\newcommand{\BV}{Brunt-V\"{a}is\"{a}l\"{a}}

\newcommand{\ddl}[2]{\frac{d \ln #1}{d \ln #2}}

\slugcomment{Accepted by The Astrophysical Journal Letters on November 8, 2016}
\accepted{November 8, 2016}

\begin{document}

\title{Convection Destroys the Core/Mantle Structure in Hybrid C/O/Ne White Dwarfs}

\author{Jared Brooks\altaffilmark{1}, Josiah Schwab\altaffilmark{2,3,4}, Lars Bildsten\altaffilmark{1,5}, Eliot Quataert\altaffilmark{2}, Bill Paxton\altaffilmark{5}}

\altaffiltext{1}{Department of Physics, University of California, Santa Barbara, CA 93106}

\altaffiltext{2}{Astronomy and Physics Departments and Theoretical Astrophysics Center, University of California, Berkeley, CA 94720, USA}
\altaffiltext{3}{Department of Astronomy and Astrophysics, University of California, Santa Cruz, CA 95064, USA}
\altaffiltext{4}{Hubble Fellow}
\altaffiltext{5}{Kavli Institute for Theoretical Physics, Santa Barbara, CA 93106}

\begin{abstract}

A hybrid C/O/Ne white dwarf (WD) -- an unburned C/O core surrounded by an O/Ne/Na mantle -- can be formed if the carbon flame is quenched in a super-AGB (SAGB) star or white dwarf merger remnant. 
We show that this segregated hybrid structure becomes unstable to rapid mixing within 2,000 years of the onset of WD cooling.    
Carbon burning includes a weak reaction that removes electrons, resulting in a lower electron-to-baryon ratio ($\Ye$) in the regions processed by carbon burning compared to the unburned C/O core, making the O/Ne mantle denser than the C/O core as the WD cools.  
This is unstable to efficient mixing.
We use the results of $\texttt{MESA}$ models with different size C/O cores to quantify the rate at which the cores mix with the mantle as they cool. In all cases, we find that the WDs undergo significant core/mantle mixing on timescales  shorter than the time available to grow the WD to the Chandrasekhar mass ($M_{\rm Ch}$) by accretion. 
As a result, hybrid WDs that reach $M_{\rm Ch}$ due to later accretion will have lower central carbon fractions than assumed thus far.
We briefly discuss the implications of these results for the possibility of Type Ia supernovae from hybrid WDs.

\end{abstract}


\section{Introduction}\label{sec:intro}

In models of $M\gtrsim 7.2 M_\odot$ SAGB stars, carbon ignites off-center, triggering a carbon flame that can in principle propagate all the way to the center, converting the core into O/Ne/Na \citep{Nomoto1985, Timmes1994, Garcia-Berro1997, Saio1998}.
However, if mixing at convective boundaries is sufficiently effective, the carbon flame may be quenched a significant distance from the core, resulting in a hybrid C/O/Ne WD (\citealt{Doherty2010, Denissenkov2013, Chen2014, Wang2014, Denissenkov2015, Farmer2015}; \citealp[see, however, ][]{Lecoanet2016}).
These hybrid C/O/Ne models consist of an unburned C/O core surrounded by a O/Ne/Na mantle.
Hybrid models have been utilized in 2D and 3D hydrodynamic codes to study the thermonuclear SNe they produce and to compare synthesized light-curve and spectra to observations of peculiar SNe Ia \citep{Kromer2015a, Willcox2016, Bravo2016}. 

However, all of these authors assumed that the compositional structure was fixed as the WD cools.
In this Letter, we show that there is significant mixing between the C/O core and the overlying O/Ne/Na mantle.
The mixing is driven by the lower electron-to-baryon ratio, $\Ye$, in the O/Ne/Na mantle that has been processed by carbon burning.   
This $\Ye$ gradient corresponds to a heavy fluid on top of a light fluid and rapidly (over $\sim$ kyrs) becomes unstable to convection as the WD cools towards an isothermal configuration.

In \S \ref{sec:stoich} we discuss the nuclear reaction in carbon burning that reduces $\Ye$.  
In \S \ref{sec:instability} we derive the criterion for a $\Ye$ gradient to be unstable to convection and compare to our numerical $\texttt{MESA}$ models.  
We discuss our $\texttt{MESA}$ models in \S \ref{sec:models} and explain our results in \S \ref{sec:results}.
We finish with our conclusions and implications in \S \ref{sec:conclusions}, highlighting that this mixing will be complete by the time any accreting WD reaches $M_{\rm Ch}$. 

\section{Carbon burning lowers $\Ye$}\label{sec:stoich}

Material processed by carbon burning will have a lower electron-to-baryon ratio, $\Ye$. 
The change in $\Ye$ has been previously discussed in the context of the carbon simmering phase in Type Ia supernova progenitors
\citep{Piro2008, Chamulak2008}.  
Carbon burning proceeds via
\begin{equation}\label{1.99}
  ^{12}\text{C}+{^{12}\text{C}} \rightarrow
  \begin{cases}
    ^{20}\text{Ne}+\alpha & (\text{0.56})\\
    ^{23}\text{Na}+p & (\text{0.44}) \hspace{2mm},
  \end{cases}
\end{equation}
where the numbers in parentheses are the branching fractions for the relevant temperatures \citep{Caughlan1988}.  
The free proton is then rapidly captured, with most captures occurring onto either $^{12}$C or $^{23}$Na.

When the proton is captured onto $^{12}$C, the reaction $^{12}$C(p,$\gamma$)$^{13}$N is followed by the beta-decay $^{13}$N($e^-$,$\nu_e$)$^{13}$C, which reduces the electron fraction. 
The $^{13}$C later captures the $\alpha$-particle generated from carbon-burning, $^{13}$C($\alpha$,n)$^{16}$O, producing a free neutron.  
This free neutron will capture onto $^{12}$C, creating $^{13}$C.  
Thus, the net result (approximating the branching fractions as equal) is that for every six $^{12}$C nuclei burned, one proton is converted into a neutron \citep{Piro2008}.  However, if the proton is captured onto $^{23}$Na, the proton-capture reaction $^{23}$Na(p,$\alpha$)$^{20}$Ne is followed only by an $\alpha$-capture.  
Since no weak reactions occur in this channel, the net result is no change in $\Ye$.

The thermally-averaged cross-section for proton captures onto $^{23}$Na is ${\approx}2$ times larger than that of $^{12}$C \citep[REACLIB; ][]{Cyburt2010}.  
At the relevant densities of $\rho\approx10^6$ g cm$^{-3}$, screening effects enhance the ratio of proton capture rates relative to those of $^{12}$C by a factor of ${\approx}1.5$. 
To demonstrate that the competition of these two proton capture channels quantitatively explains our more detailed \texttt{MESA} results, we use a simple model that assumes equal branching fractions in equation \eqref{1.99} and assumes that when $X_{\rm C12} > 3 X_{\rm ^{23}Na}$ all protons capture on to $^{12}$C and when $X_{\rm C12} < 3 X_{\rm ^{23}Na}$ all protons capture on to $^{23}$Na (meaning its consumption rate exactly matches its production rate).
For complete carbon burning, this simple model gives values for the $^{23}$Na abundance and total neutronization that agree with our $\texttt{MESA}$ models. 
This results in a factor of ${\approx}2$ less neutronization than suggested by the counting arguments of \cite{Piro2008}, because approximately half of the protons produced in equation \eqref{1.99} are captured by $^{12}$C and half by $^{23}$Na when the burning is complete.
Therefore, a change in the electron-to-baryon ratio due to complete burning is
\begin{equation}
  \label{1.6}
  \Delta \Ye \approx 0.5 \left(\dfrac{\Delta X(^{12}\text{C})}{72}\right)\hspace{2mm}. \end{equation}
In our models, the initial carbon mass fraction is $0.345$ and the carbon burns completely.  Consistent with equation~\eqref{1.6}, the difference in $\Ye$ between burned and unburned material is $\Delta Y_e=-0.00235$.

\section{Convective Instability Criterion}\label{sec:instability}

The reduction in $\Ye$ due to carbon burning creates a sharp $\Ye$ gradient between burned and unburned material. 
Since the ashes above are much hotter than the unburned material below, this configuration is stable against convection.  
However, if the flame is extinguished before reaching the center, the stabilizing thermal gradient vanishes under the action of neutrino cooling and electron conduction, while the $\Ye$ gradient remains.
By coincidence, the $\Delta \Ye/\Ye$ is the same order of magnitude as the
ratio of thermal energy to the Fermi energy.  
Performing a simple buoyancy calculation by adiabatically perturbing a fluid element in an isothermal configuration of a degenerate, non-relativistic electron gas, and an ideal ion gas with average ion charge $\zbar$ gives the convective instability criterion as
\begin{equation}\label{eq:old-stability} \dfrac{d\ln \Ye}{d\ln P} > \dfrac{k_BT}{\gamma\bar{Z}E_F} \hspace{2mm} ,
\end{equation}
where $E_F$ is the electron Fermi energy, and $\gamma$ is the adiabatic exponent.
Therefore, as the WD cools, the region with a $\Ye$ gradient
is destined to convect.

A more accurate convective stability criterion can be determined from the \BV\ frequency, which is 
\begin{equation}
  N^2 = \frac{g^2 \rho}{P}\frac{\chiT}{\chir}
  \left(\grada-\gradT+B\right)
  \label{eq:brunt_B}
\end{equation}
where $\gradT$ and $\grada$ are the actual and adiabatic temperature gradients with respect to the pressure, $B$ is the term that takes into account the effect of composition gradients \citep[see eq.~6 in][]{Paxton2013}, $g$ is the local gravitational acceleration, $\rho$ and $P$ are the density and total pressure, respectively, and $\chi_Q=\partial\ln P/\partial\ln Q$, where the choices for $Q$ are $\rho, T, \abar$, or $\zbar$, and the remaining three variables are held fixed.  
For an equation of state where the influence of the composition can be fully specified by the mean ion weight, $\abar$, and the mean ion charge, $\zbar$,
\begin{equation}\label{new_B}
  B = -\frac{1}{\chiT} \left[
    \chi_{\abar}\frac{\dlna}{\dlnp} +
    \chi_{\zbar}\frac{\dlnz}{\dlnp}
  \right] ~.
\end{equation}
In order to analytically evaluate the partial derivatives in equation~\eqref{new_B}, we assume a simple form for the equation of state consisting of fully-degenerate electrons, an ideal gas of ions, and the Coulomb pressure in the Wigner-Seitz approximation.  
That is, we take $\ptot = \pele + \pion + \pcou$, where $\pele = P_0 \left(\rho\Ye/\rho_0\right)^\gamma$, $\pion = \rho k_B T/(m_p \abar)$, and $\pcou = -3 \Gamma \pion/ 10$.  
The Coulomb coupling parameter $\Gamma = (\zbar e)^2/(a_i k_B T)$ where $a_i = (4 \pi \rho/3 m_p\abar)^{-1/3}$. 

Plugging in these definitions, equations~\eqref{eq:brunt_B} and \eqref{new_B} imply that the gradient in electron fraction corresponding to the neutral stability condition, $N^2 = 0$, is

\begin{equation}
  \frac{\dlnye}{\dlnp} = \frac{\chiT \left(\grada - \gradT \right)}
  { \gamma \left(1-\epsilon+\frac{3}{10}\Gamma\epsilon\right)
        - \frac{3}{5}\Gamma\epsilon - \left(1+\frac{1}{5}\Gamma\right)\epsilon \ddl{\abar}{\Ye}}~.
  \label{eq:new-stability}
\end{equation}
where for convenience we define $\epsilon = \pion/\ptot$.  
Gradients steeper than equation \eqref{eq:new-stability} are unstable to convection.
Note that the critical gradient corresponding to $N^2=0$ evolves as $\gradT$ evolves within the WD.  
We use equation \eqref{eq:new-stability} to compute the neutral stability lines shown in \S \ref{sec:models} and \S \ref{sec:results}.

\section{Rapid Onset and Modeling of Convection}\label{sec:models}

We generated models of C/O/Ne hybrid WDs using the published set of controls from \cite{Farmer2015} and $\texttt{MESA}$ version 8118 \citep{Paxton2011, Paxton2013, Paxton2015a}.
These models have C/O cores of mass $0.24, 0.40,$ and $0.55 M_\odot$ and total masses of $1.12, 1.09,$ and $1.07 M_\odot$, respectively.
The stars producing these WDs had zero-age main sequence masses of $7.2-7.5 M_\odot$.
The H rich envelopes are removed once the inward-moving carbon flame dies, simulating a thermally pulsing AGB phase, or interaction with a binary companion.
We then restart these models with all the same controls except all forms of overshooting and thermohaline mixing turned off.

\begin{figure}[h]
  \centering
  \includegraphics[width = \columnwidth]{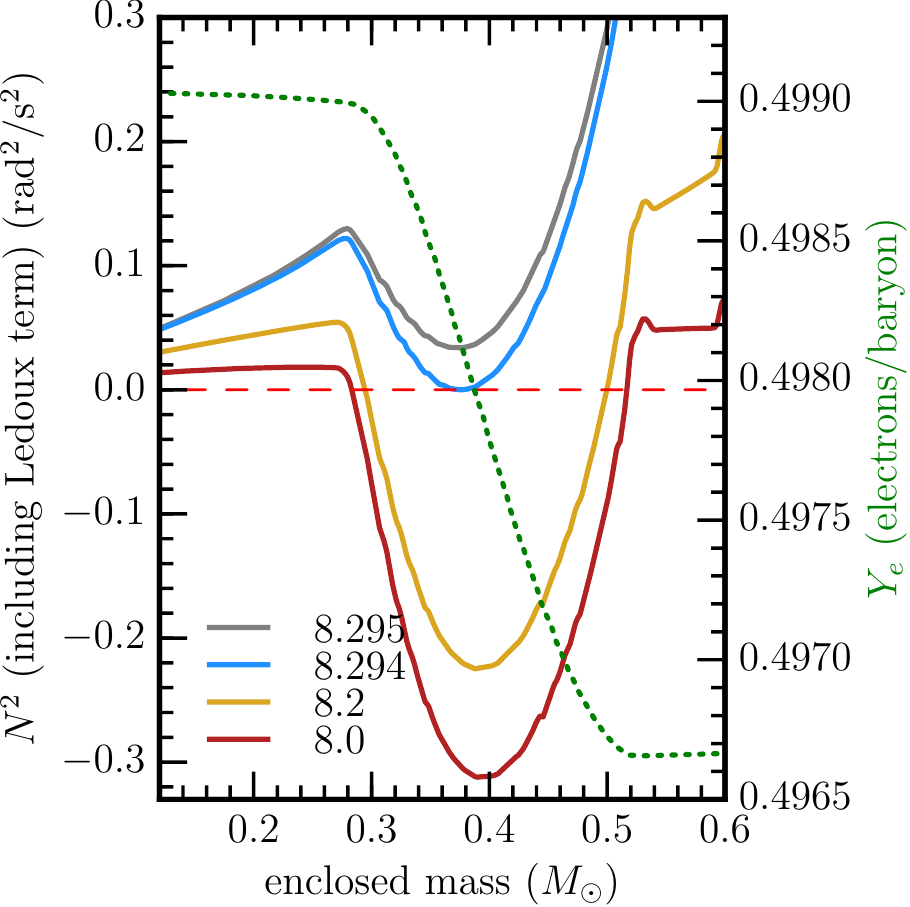}
  \caption{\footnotesize The Brunt-V$\ddot{\rm a}$is$\ddot{\rm a}$l$\ddot{\rm a}$ frequency squared (left y-axis) and $\Ye$ profile (right y-axis) for the $\texttt{MESA}$ model with a $0.40 M_\odot$ C/O core ($1.09 M_\odot$ total mass) in which the Schwarzschild criterion for convection is used, so convection driven by the $\Ye$ profile is inhibited.
  The plotted Brunt-V$\ddot{\rm a}$is$\ddot{\rm a}$l$\ddot{\rm a}$ frequency includes the Ledoux term.
  The profiles are labelled by the log of their central temperatures.
  As the temperature decreases and the density increases, $N^2$ becomes more negative, and therefore more unstable to convection.
  The blue curve at $\log T_c=8.294$ is the time at which convection would begin, only 1600 yrs after the formation of the hybrid WD.
  The cooling ages to $\log T_c=8.2, 8.0$ are 13.5 kyr and 330 kyr, respectively.}
  \label{fig:1}
\end{figure}

We started by running a $\texttt{MESA}$ simulation with the Schwarzschild criterion for convection, which ignores compositional gradients, to show that the square of the total Brunt-V$\ddot{\rm a}$is$\ddot{\rm a}$l$\ddot{\rm a}$ frequency (equation \ref{eq:brunt_B}) becomes negative as the WD cools (see Figure \ref{fig:1}).
The regions that become unstable to convection ($0.28-0.52 M_\odot$) coincide with the steepest portion of the $\Ye$ profile, shown by the dotted green curve in Figure \ref{fig:1}.
Cooling erases the stabilizing temperature gradient, and hence mixing will begin, within a mere 1.6 kyr of the formation of the hybrid WD, as shown by the blue profile in Figure \ref{fig:1} at $\log T_c=8.294$.
The cooling is dominated by neutrino emission until $\log T_c\approx 7.8$.

Figure \ref{fig:1} highlights that the WD rapidly evolves towards convective instability, which can in principle mix material on the much shorter dynamical timescale. 
Given that this dynamical mixing timescale is so much faster than the timescale on which the WD evolves due to cooling, we expect that convection will (as usual) lead to nearly zero Brunt-V$\ddot{\rm a}$is$\ddot{\rm a}$l$\ddot{\rm a}$ frequency.
Thus, for our assumed thermal evolution, we expect convection will mix the composition efficiently enough to keep the actual $\Ye$ gradient extremely close to neutral stability at all times.  

Evolving the hybrid WD models using the standard MESA MLT implementation leads to severe numerical difficulties.  
When taking a timestep, as the \texttt{MESA} solver iterates to find the next model, some zones alternate between being convective and non-convective.  
This corresponds to the temperature gradient ($\nabla_T$) at cell faces switching between the adiabatic temperature gradient ($\nabla_{\rm ad}$) and the radiative temperature gradient ($\nabla_{\rm rad}$).  
These are significantly different, since the temperature profile is highly stable against convection and only being destabilized by the $\Ye$ gradient.  
As a result, the solver fails to find models that satisfy its ``\texttt{dlnTdm}'' equation (eq.~8 of \citealt{Paxton2011}). 
In order to circumvent this issue, we set the $\texttt{MESA}$ control $\texttt{mlt\_gradT\_fraction = 0}$, which sets $\nabla_T$ in convective regions to $\nabla_{\rm rad}$ instead of $\nabla_{\rm ad}$.  
The choice to set the temperature gradient to the radiative gradient in convection zones is appropriate when the convection is thermally inefficient and the energy flux is dominated by conduction.   
Thus our choice assumes that the convection does not significantly modify the thermal evolution of the WD.

\texttt{MESA} treats mixing, including convective mixing, as a diffusive process \citep{Paxton2011}. 
When using standard MLT, \texttt{MESA} sets the diffusion coefficient to be $\tfrac{1}{3}v_{\mathrm{c}} \Lambda$, where $v_{\mathrm{c}}$ is the convective velocity and $\Lambda$ is the mixing length. 
Our models consistently exhibit convective regions that span only a few cells and thus are much smaller than a pressure scale height.  
Using a mixing length of order the pressure scale height likely overestimates the mixing, but limiting the mixing length to the size of the convective region introduces an undesired dependence of the mixing on the numerical resolution.  
In order to circumvent this issue, we set the diffusive mixing coefficient in convectively unstable zones to be proportional to the thermal diffusion coefficient, that is $D_{\rm mix}=\beta D_{\rm thermal}$ with $\beta$ a constant.  
The choice of $D_{\rm thermal}$ as the reference diffusivity allows us to ensure that the mixing remains more rapid than the thermal evolution. 
Physically, we expect values of $\beta \gg 1$ to approach the correct answer, representing the fact that convection can mix material on timescales much shorter than the thermal timescale (which is the timescale on which the convective regions are created; see Fig.~\ref{fig:1}).

\begin{figure}[h]
  \centering
  \includegraphics[width = \columnwidth]{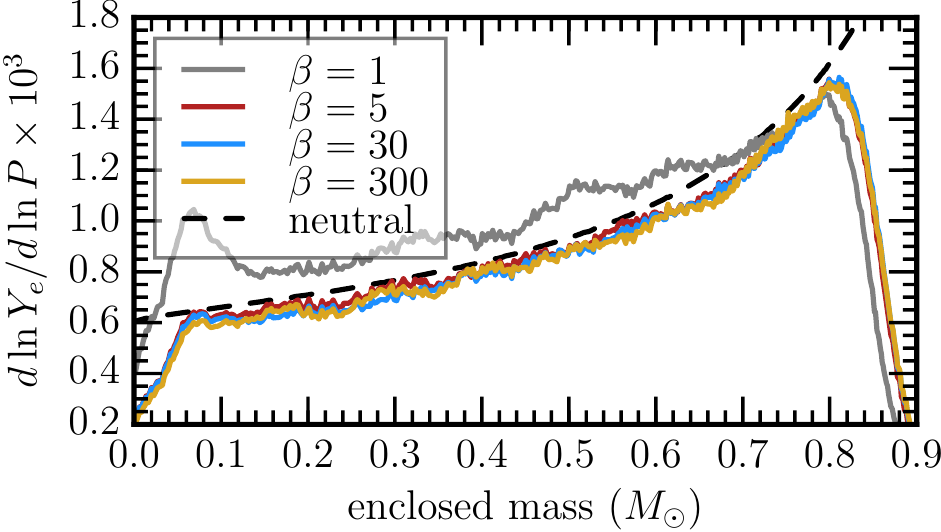}
  \caption{\footnotesize Profiles of $d \ln Y_e/d \ln P$ for $\log T_c=8.2$ , compared to the neutrally buoyant $\Ye$ gradient (black dashed curve, eq. \ref{eq:new-stability}) for the $\texttt{MESA}$ model with a $0.40 M_\odot$ C/O core ($1.09 M_\odot$ total mass).
  The solid curves show profiles for $\texttt{MESA}$ models with the full Ledoux convection criterion and different choices of $\beta=D_{\rm mix}/D_{\rm thermal}$ (see \S \ref{sec:models}).
  The profile for $\beta=1$ underestimates mixing compared to the neutral buoyancy curve.
  The profiles for $\beta=5, 30, 300$ are similar, and consistent with the expected neutral stability curve.  
  }
  \label{fig:3}
\end{figure}

We compare the resulting composition profiles for different choices of $\beta$ in Figure \ref{fig:3}.
While the choice of $\beta=1$ underestimates the mixing compared to neutral buoyancy as expressed by equation \eqref{eq:new-stability}, shown by the grey curve in Figure \ref{fig:3}, the choices of $\beta=5, 30,$ and $300$ yield nearly identical results and reach the neutral buoyancy condition derived in \S \ref{sec:instability}.   
This supports our claim that the $\Ye$ gradients should match those of neutral stability (eq.~\ref{eq:new-stability}), as long as the mixing timescale is shorter than the thermal timescale.   
We also found comparable results using alternative ways of controlling the mixing in $\texttt{MESA}$, such as using a mixing length alpha parameter that varied in proportion to the thermal timescale of the WD.   
The rest of our results  employ $D_{\rm mix}=\beta D_{\rm thermal}$ with $\beta=5$.

\section{Results for a range of Core Masses and Thermohaline Mixing}\label{sec:results}

\begin{figure}[h]
  \centering
  \includegraphics[width = \columnwidth]{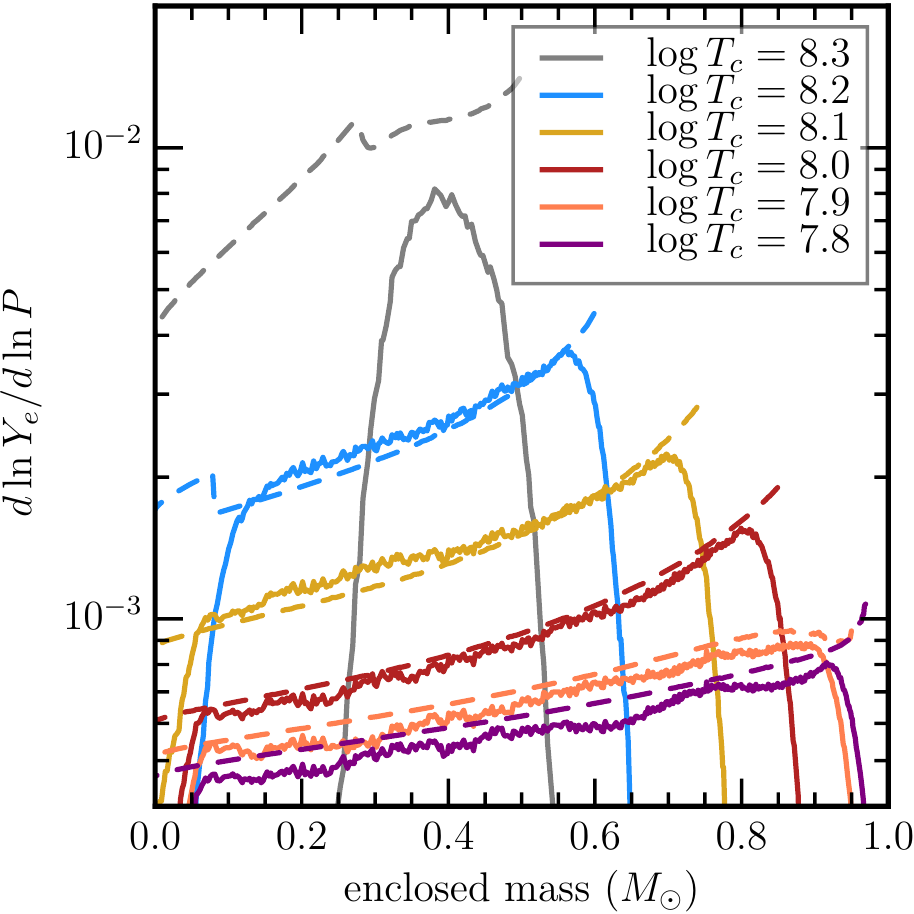}
  \caption{\footnotesize   Profiles of $d \ln Y_e/d \ln P$ as a function of time (labeled by the central temperature) for the $\texttt{MESA}$ model with an initial C/O core mass of $0.4 M_\odot$ (solid curves), compared to the neutrally buoyant $\Ye$ gradient (dashed curve, eq. \ref{eq:new-stability}).    
  The dashed neutral buoyancy lines are cropped to show only the regions where convective mixing has occurred. 
  As the WD cools and becomes isothermal, the core becomes increasingly mixed by convection driven by the unstable composition gradient.}
  \label{fig:2}
\end{figure}

\begin{figure}[h]
  \centering
  \includegraphics[width = \columnwidth]{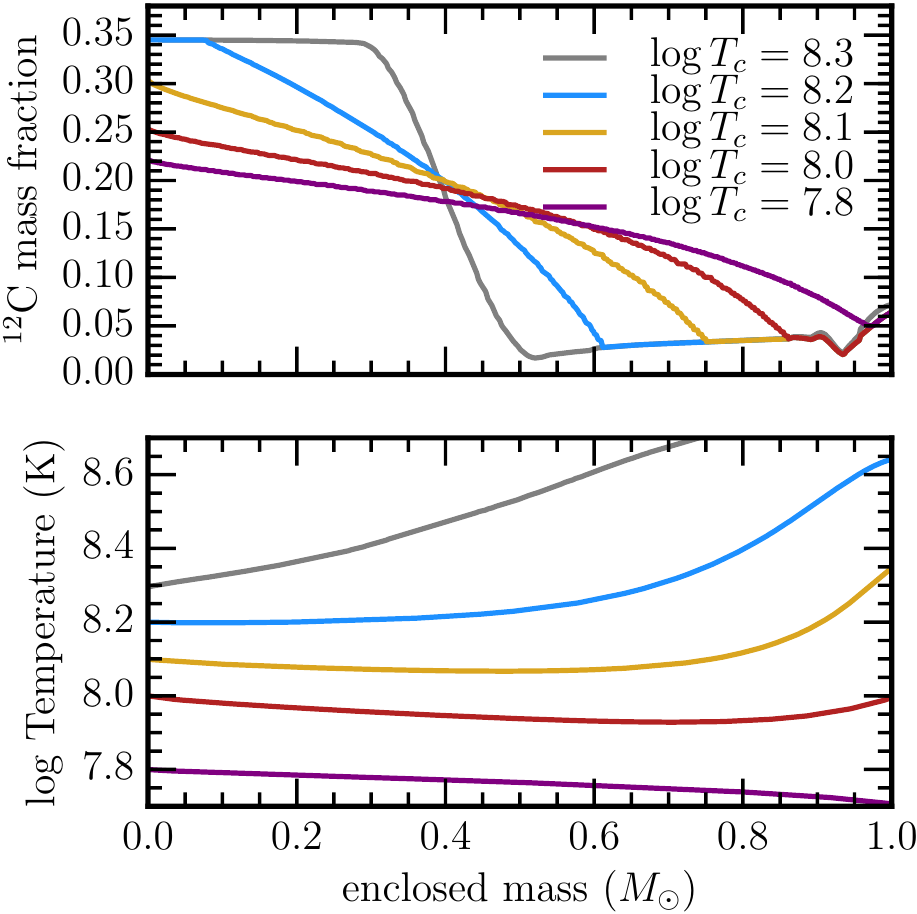}
  \caption{\footnotesize  Top panel: Carbon mass fractions as a function of time (labeled by central temperature), for the $\texttt{MESA}$ model with an initial C/O core mass of $0.4 M_\odot$.   
  As the WD cools and becomes isothermal, the core becomes increasingly well mixed and the central carbon fraction decreases.  
  Bottom panel: Temperature profiles at the same time as the profiles in the top panel.
  }
  
  \label{fig:4}
\end{figure}

Figure \ref{fig:2} shows the composition evolution of our $\texttt{MESA}$ model with an initial $0.4 M_\odot$ C/O core including the effects of convective mixing during the subsequent thermal evolution (but not including thermohaline mixing).   As the WD cools, the  $\Ye$ gradients (solid lines) become shallower and shallower, comparable to what we analytically expect given the  the neutral buoyancy gradients (dashed lines).    Figure \ref{fig:4} shows the carbon mass fractions at these same times, demonstrating that the C/O core and O/Ne/Na mantle rapidly mix as the WD cools.

\begin{figure}[h]
  \centering
  \includegraphics[width = \columnwidth]{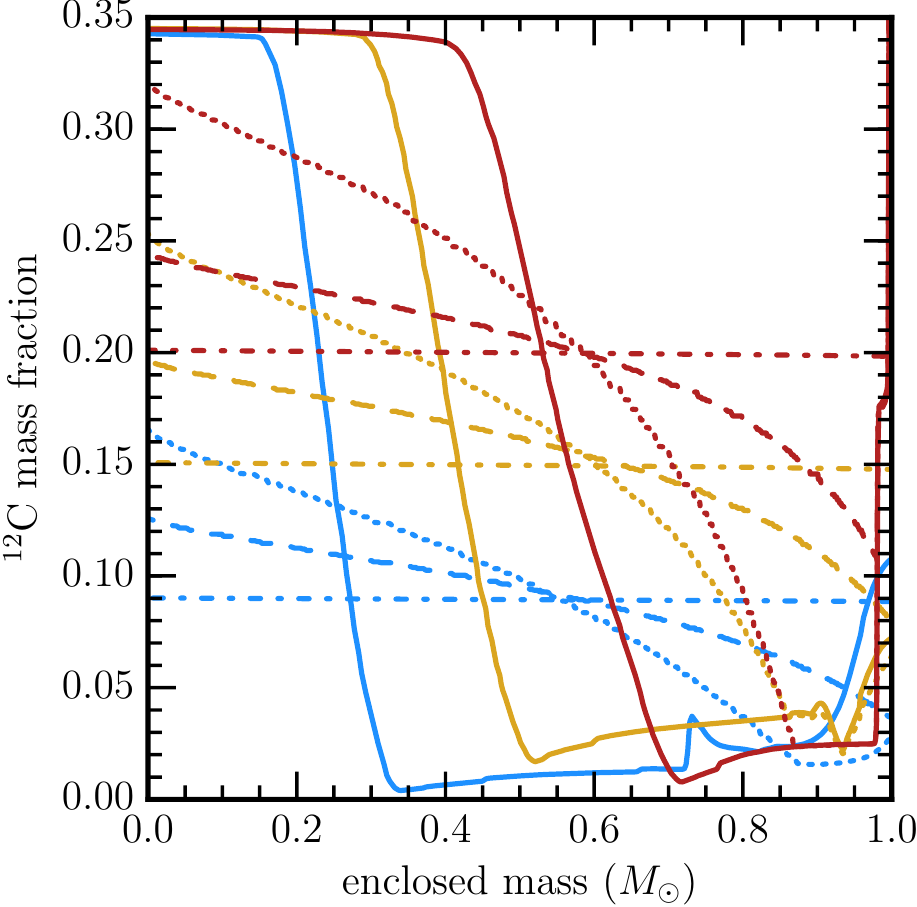}
  \caption{\footnotesize   Carbon mass fractions as a function of time (labeled by central temperature).   
  Red curves are for a hybrid WD with an initial $0.55 M_\odot$ C/O core, orange curves a $0.40 M_\odot$ C/O core, and blue curves a $0.24 M_\odot$ C/O core.
  Solid curves show the carbon mass fraction before any mixing occurs,  dotted when $\log T_c=8.0$, and  dashed when $\log T_c=7.5$; the dot-dashed curve also at $\log T_c=7.5$ includes thermohaline mixing.
  }
  \label{fig:5}
\end{figure}

Figure \ref{fig:5} shows the $^{12}$C mass fraction profiles of three models with different initial C/O core masses before any mixing (solid curves), when $\log T_c=7.8$ (dotted curves), and when $\log T_c=7.5$ (dashed curves).

All of the results discussed above are based on models run without including the effects of thermohaline mixing.
As the WD evolves towards an approximately isothermal profile, mixing by convection alone would leave behind a finite $\Ye$ gradient given by setting $\nabla_T=0$ in equation \eqref{eq:new-stability}. 
However, such a composition gradient is unstable to thermohaline convection.
Therefore, additional mixing will continue to occur.   Prescriptions for thermohaline mixing \citep{Kippenhahn1980, Brown2013} remain somewhat uncertain, however, leading to more substantial uncertainties in the resulting composition profile.
As illustrative examples, we carried out $\texttt{MESA}$ calculations with $\texttt{thermohaline\_option = `Kippenhahn'}$ and $\texttt{thermohaline\_coeff = 1}$.
This diffusion coefficient proposed by \cite{Kippenhahn1980} as written in eq. (14) of \cite{Paxton2013}, with a dimensionless, multiplicative efficiency factor of $\texttt{thermohaline\_coeff}$, which we set to 1.

Figure \ref{fig:5} includes the $^{12}$C mass fraction profiles from all three models run with thermohaline mixing, at the time when $\log T_c=7.5$ (dot-dashed lines).   
In this implementation, thermohaline mixing leads to fully mixed WDs at this time, with completely flat  $^{12}$C mass fraction profiles.    
Given the uncertainties in thermohaline mixing, however, the dashed (no thermohaline mixing) and dot-dashed (efficient thermohaline mixing) models in Figure \ref{fig:5} likely bracket the  mass fraction profiles at this time.

\section{Conclusions}\label{sec:conclusions}

We have shown that hybrid C/O/Ne WDs created from quenched carbon flames\footnote{A recent multi-D study of the convectively-bounded carbon flames \citep{Lecoanet2016} suggests that the bouyancy barrier accross the flame is too great to permit sufficient mixing to quench the flame, implying that hybrid C/O/Ne WDs are unlikely to form in the first place.} are unstable to convective mixing as the WD cools. 
Weak reactions during carbon burning cause the O/Ne ash to have a lower $\Ye$ than the C/O fuel. 
The configuration of a heavy O/Ne mantle above the cold C/O core is initially stable due to the higher temperatures of the freshly burned O/Ne mantle. 
However, as neutrino cooling and electron conduction reduce the temperature, the hybrid configuration becomes convectively unstable.    
Convective mixing rapidly leads to a much more uniform composition profile (Fig. \ref{fig:5}).

Our modeling choices for this initial work lead to a profile that is very nearly neutrally buoyant (Figure \ref{fig:2}).
We did not explore the physics of thermohaline convection for this neutrino-cooled material, which remains an open question.
However, at a minimum, neutral buoyancy would always be reached.
We also do not allow the resulting convection to modify the thermal evolution of the WD (\S \ref{sec:models}).
This effect may be significant at the earliest times.
Future work can address these questions and their impact on the compositional structure of hybrid WDs.

To understand the importance of this mixing for possible Type Ia scenarios, it is useful to compare the mixing times to the timescales for hybrid WDs to gain mass and approach the Chandrasekhar mass $M_{\rm Ch}$.
The cooling times to the central temperatures of $\log T_c = 8.2, 8.0, 7.8$, and $7.5$ shown in Figures \ref{fig:2} \& \ref{fig:4} are $13.5$ kyr, $330$ kyr, $6.4$ Myr, and $77$ Myr, respectively.
For comparison, the time necessary to grow a $1.0 M_\odot$ WD to $M_{\rm Ch}$ at steady H and He burning rates is approximately $2$ and $0.2$ Myr, respectively.
Furthermore, this does not take into account the time between WD formation and the start of mass transfer, which is often much longer than the growth time via accretion.
Therefore, hybrid WDs in binary scenarios are likely to be even more mixed once they reach $M_{\rm Ch}$.   
Existing models of Type Ia SNe from hybrid WDs neglect this compositional mixing \citep{Denissenkov2015, Kromer2015a,Bravo2016}.

If a mixed hybrid core can grow to $M_{\rm Ch}$, it is unclear whether its fate will be similar to that found in existing calculations of hybrid C/O/Ne WD explosions as candidate progenitors of some Type Ia SNe.
When the core and mantle are fully mixed, the central carbon fraction is reduced significantly (see Figure \ref{fig:5}). 
This implies that higher densities must be reached before  carbon ignition and the subsequent carbon simmering phase can begin.  
The additional $^{23}$Na and $^{25}$Mg mixed into the core will increase the amount of Urca-process cooling that occurs \citep{Denissenkov2015, Martinez-Rodriguez2016}, potentially delaying carbon ignition to yet higher densities.   
Higher density ignition of carbon will change the nucleosynthetic signature of  any resulting explosion \citep{Townsley2009, Townsley2016}.  
In addition, the higher densities required for carbon ignition at low central carbon fractions can lead to exothermic electron-capture reactions on  $^{24}$Mg and $^{20}$Ne setting in before  carbon ignition densities are reached.   
This would give rise to qualitatively different evolution, potentially leading to accretion induced collapse of the white dwarf to a neutron star rather than a convective-core carbon burning runaway (e.g., \citealt{Schwab2015a}).

\

We  thank Frank Timmes for useful conversations on nuclear reaction rates involving carbon.
This research is funded in part by the Gordon and Betty Moore Foundation through Grant GBMF5076 to L.B. and E.Q..
We acknowledge stimulating workshops at Sky House where these ideas germinated. 
This work was supported by the National Science Foundation under grants PHY 11-25915, AST 11-09174, and AST 12-05574.   E.Q. was supported in part by a Simons Investigator Award from the Simons Foundation and the David and Lucile Packard Foundation.    J.S. was supported by the National Science Foundation Graduate Research Fellowship Program under Grant No. DGE 11-06400 and by NSF Grant No. AST 12-05732. 
Support for this work was provided by NASA
through Hubble Fellowship grant \# HST-HF2-51382.001-A awarded by the
Space Telescope Science Institute, which is operated by the
Association of Universities for Research in Astronomy, Inc., for NASA,
under contract NAS5-26555.

\bibliographystyle{aasjournal}

\end{document}